\documentclass{article}
\RequirePackage{amsmath} % Load amsmath before llncs document class

\usepackage[compatibility=false]{caption}

\usepackage{booktabs} % For pretty tables
\usepackage{subcaption} % For sub-figures
\usepackage{graphicx}
\usepackage{pgfplots}
\usepackage[all]{nowidow}
\usepackage[utf8]{inputenc}
\usepackage{tikz}
\usetikzlibrary{er,positioning,bayesnet}
\usepackage{multicol}
\usepackage{algpseudocode,algorithm,algorithmicx}

\usepackage{hyperref}
\usepackage[inline]{enumitem} % Horizontal lists

\definecolor{blue}{HTML}{1F77B4}
\definecolor{orange}{HTML}{FF7F0E}
\definecolor{green}{HTML}{2CA02C}

\pgfplotsset{compat=1.14}

\usepackage[numbers,round]{natbib} % Use natbib citation style with numerical citations
\begin{document}
\title{Determinants of Uruguay's Real Effective Exchange Rate: A Mundell-Fleming Model Approach}
\author{Didarul Islam,  Mohammad Abdullah Al Faisal}

\maketitle              % typeset the header of the contribution
\begin{abstract}
This study examines the factors influencing the short-term real effective exchange rate (REER) in Uruguay by applying an extended Mundell-Fleming model. Analyzing the impact of the US lending rate (USLR), money supply (M2), inflation (CPI), and the world interest rate (WIR), the paper uses a linear regression model with Newey-West standard errors. Key findings reveal that an increase in the USLR, CPI, and M2 is associated with a depreciation of the REER. In contrast, WIR shows no significant impact. These findings are consistent with the theoretical expectations of the Mundell-Fleming model regarding open economies under floating exchange rates. Therefore, authorities should tighten monetary policy, control inflation, adjust fiscal strategies, and boost exports in response to Peso depreciation.

\end{abstract}

\section{Introduction}

Uruguay is a small open economy with its regional dynamics, making it a relevant case study for analyzing the determinants of real effective exchange rates (REER) in the context of the Mundell-Fleming model. Having recovered from the severe financial crisis of 2002, intricately linked to the economic turmoil in neighboring Argentina and broader regional events, Uruguay has navigated significant challenges. The crisis not only destabilized its banking system and currency but also had implications on its fiscal policy. It eventually led to the fluctuation of Uruguay’s REER, given its heavy reliance on a dollarized banking system and public debt. Post-crisis, Uruguay has exhibited economic robustness, experiencing a notable recovery since the second quarter of 2003, evidenced by renewed investor confidence and growth in sectors such as agriculture, tourism, and technology. This economic growth, marked by a more competitive REER and supported by institutional credibility, has opened new avenues for growth. The changing nature of Uruguay’s REER, especially in the context of regional economic trends with Argentina and Brazil, has had significant implications for its trade balance, economic stability, and growth path. Understanding these dynamics, mainly the interaction between domestic economic policies and external regional factors, is crucial for formulating Uruguay’s economic strategy and policy response \citep{hausmann2005}.

During the onset of the United States recession in 2008, capital outflowed from developed countries to emerging countries, prompting advanced economies to adopt both conventional measures like monetary easing and unorthodox policies such as the purchase of long-term assets \citep{kolasa2020}. For dealing with such shocks in developing economies, it is argued that interventions in foreign exchange are highly useful \citep{blanchard2015,daude2016,bucacos2019}. However, Al FAISAL and Islam \citep{alfaisal2022} argue that the effects of these shocks depend on the receptive capacity of the recipient economies. In response to these capital flow shocks, Uruguay adopted the sterilized foreign exchange along with an asset and liability approach \citep{vicente2017}. Bucacos et al. \citep{bucacos2019} claimed that these two policies improved the foreign exchange market for Uruguay.

Since Uruguay operates under a floating exchange rate, it has witnessed fluctuations in Peso value against the U.S. dollar. According to the Economic Survey of Latin America and the Caribbean 2023, Peso appreciated in 2022 and it continued appreciating until the beginning of 2023, but, at the end of the first half of 2023, its value returned to its previous status. In the meantime, almost no actions were taken by the central bank of Uruguay. However, the situation demands some policies to be adopted to improve the present status quo. 

Since the determinants of REER were not investigated before, this paper aims to contribute to this field of study in the context of the Mundell-Fleming model and analyze key economic indicators that shape the dynamics of exchange rates of Uruguay in relation to an interconnected world economy.
In the first section, the study introduces some specific ongoing problems of the Uruguayan economy and the necessity of this paper to resolve those difficulties. Section 2 recaps the national and international literature on this topic. Section 3 discusses the nature of the data, and section 4 builds a model and explains its empirical results and the results of diagnostic tests. Finally, section 5 offers a conclusion with some policy recommendations.

\section{Literature Review}
Many scholars extended the Mundell-Fleming model in different ways, and many investigated the validity of Mundell-Flemming in numerous settings. Depending on the nature of economies or studies, they found different results. Some align with the theoretical expectations, and some do not. Even though, to the best of researchers’ knowledge, the determinants of Uruguay’s real effective exchange rate were not examined before, we summarize some similar national works and some relevant international studies in this section. 

Mordecki and Miranda \citep{mordecki2019} checked the Real Exchange Rate (RER) volatility by employing GARCH and IGARCH for Uruguay, Brazil, New Zealand, and Chile spanning 1990 to 2013. They tried to find the effect of RER volatility on total exports by applying Johansen’s methodology. They found that global demand and international prices affect exports positively in all the economies except Uruguay. 

Examining Exchange Rate Pass Through (ERPT) in Uruguay and Chile, Cuitiño et al. \citep{cuitino2022} found a significant degree of imperfect credibility in Uruguay's monetary policy, combined with the crucial role of exchange rate stabilization. The estimations stressed the interconnected nature of these factors, recommending a simultaneous approach by Uruguay's monetary authority. In contrast, Chile exhibited firmed inflation expectations and minimal exchange rate stabilization, presenting a distinct policy environment.

Employing a new method and using data from 1955 to 2019, Benelli \citep{benelli2023} found that exchange rate overvaluation created an opportunity for accumulating capital causing the economy of Uruguay to move from an agricultural economy to an industrial one. Capital accumulation continued precipitating the industrial sector to grow and making the Uruguayan economy highly dependent on it, but this process was favoring to keep the exchange high. 

By using the structural VAR model, Drine and Rault \citep{drine2004} investigated the causes of RER fluctuations for Uruguay, Morocco, and The Philippines through variance decomposition and impulse response function. They found that most of the fluctuations were explained by domestic shocks, and the effectiveness of printing money to increase competitiveness was weak. They also claimed that the foreign interest rate and the terms of exchange triggered RER fluctuations. 

Giannellis and Koukouritakis \citep{giannellis2013} studied the relationship between inflation rate and currency undervaluation for Uruguay, Brazil, Venezuela, and Mexico by using a non-linear model. They found that the inflation rate persisted when the domestic currency highly depreciated, and it was transitory when the depreciation of the domestic currency was slower or stable.

Licandro \citep{licandro2000} suggested adopting inflation targeting, with the exchange rate as the primary policy instrument considering the present status quo of the Uruguayan economy. 

Utilizing the Granger causality test and impulse response function of the VAR model, Asogwa et al. \citep{asogwa2016} checked the validity of the Mundell-Fleming model for the Nigerian economy by collecting data from the Central Bank of Nigeria from 1970 to 2012. While the impulse response function indicated that the Mundell-Fleming model held for the Nigerian economy, Granger causality exhibited that there was only a one-directional causality running from net export to foreign direct investment. 

Gharaibeh \citep{gharaibeh2017} aimed to find the main determinants of the real exchange rate (RER) in Bahrain using OLS regression. He found that though the effect of trade structure, balance of payment, gross domestic product, and money supply was positive on RER, foreign investment outflow, inflation, and interest rate negatively affected the RER of Bahrain.

To infer from the time-series data spanning 1981 to 2012, Orji \citep{orji2015} used the Error Correction Model (ECM) to search for the determinants of RER for Nigeria. He found that the main sources of RER fluctuations were oil revenue and interest rate differential, not the productivity differential.

In studying the nature of the exchange rate in the short term, Hsing \citep{hsing2006} found that RER is positively related to deficit spending and negatively associated with the world interest rate, country risk, real M2, and the expected inflation rate in the economy of Venezuela.

The findings from Islam’s paper confirmed that in Bangladesh, the exchange rate (ER) plays a crucial role in influencing the long-term flow of credit from banks to the private sector \citep{islam2022}.

Each of these studies has provided insights into different aspects of Uruguay’s exchange rate dynamics. However, no existing research has comprehensively explored what drives Uruguay’s exchange rate specifically through the lens of the extended Mundell-Fleming model. In the context of the Mundell-Fleming Model, the IS curve illustrates the equilibrium where total spending (consumption plus investment) equals total output, while the LM curve represents the point where the demand for money equals the supply. The Mundell-Fleming model extends the IS-LM framework to open economies, introducing the balance of payments and exchange rates into the analysis. The study’s relevance in the context of the Mundell-Fleming model lies in its empirical analysis of the determinant variables and their impact on the REER in Uruguay \citep{mundell2001}. This analysis not only supports the theoretical predictions of the model but also provides practical insights for policymakers in managing an open economy like Uruguay’s under floating exchange rates. Such insights are crucial for formulating appropriate monetary and fiscal policies to achieve desired economic outcomes, like exchange rate stability and sustainable growth.

In summary, while various studies have examined different facets of Uruguay’s exchange rate dynamics, none have comprehensively investigated the determinants of its real effective exchange rate within the framework of the Mundell-Fleming model. This study seeks to address this gap by empirically analyzing key economic indicators and their impact on Uruguay’s REER, thereby contributing to a deeper understanding of the country’s exchange rate dynamics and informing policy decisions.

\section{Data Description}

The dataset for this study is derived from the International Monetary Statistics by the IMF and the Federal Reserve Economic Data. It spans from 2001.Q4 to 2021.Q3 and includes the variables listed in Table \ref{tab:my-table}.

\begin{table}[ht]
\centering
\caption{Summary of Dataset Variables}
\label{tab:my-table}
\begin{tabular}{@{}lllll@{}}
\toprule
Variable & Source & Minimum & Mean & Maximum \\ \midrule
REER (Real Effective Exchange Rate) & IMF & 69.27 & 98.21 & 118.62 \\
USLR (US Lending Rate) & IMF & 3.250 & 4.412 & 8.250 \\
M2 (log Money Supply) & IMF & 25.61 & 26.75 & 28.04 \\
CPI (Inflation) & IMF & 3.507 & 8.699 & 27.912 \\
WIR (World Interest Rate) & Federal Reserve & 0.6506 & 3.0369 & 5.1061 \\ \bottomrule
\end{tabular}
\end{table}

M2, the log of broad money in domestic currency, reflects the money supply within Uruguay (Hsing, 2006). It encompasses cash, checking deposits, and easily convertible near money. Following Hsing (2006), the 10-year U.S. Treasury bond yield has been used as an indicator of global interest rate trends. CPI, the Consumer Price Index for all items, measures the percentage change in prices period-to-period and serves as an inflation indicator.

\begin{table}[ht]
\centering
\caption{Augmented Dickey-Fuller Test Results}
\label{table:adf_test_results}
\begin{tabular}{lccc}
\hline
Variable                 & Test Statistic & Critical Value (5\%) & Significant \\
\hline
REER                     & 0.602          & -1.95                &             \\
USLR                     & -0.859         & -1.95                &             \\
M2                       & 3.341          & -1.95                &            \\
CPI                     & -2.143         & -1.95                &         \\
WIR                      & -1.306         & -1.95                &             \\
REER (First Difference)  & -4.837         & -1.95                & ***         \\
USLR (First Difference)  & -3.265         & -1.95                & **          \\
M2 (First Difference)    & -2.203         & -1.95                & **          \\
CPI (First Difference)  & -5.183         & -1.95                & ***         \\
WIR (First Difference)   & -3.843         & -1.95                & ***         \\
\hline
\end{tabular}
\end{table}

Table \ref{table:adf_test_results} presents the results of Augmented Dickey-Fuller tests conducted on Real Effective Exchange Rate (REER), US Lending Rate (USLR), Money Supply (M2), Consumer Price Index (CPI), and World Interest Rate (WIR), along with their first differences. The test statistics are compared against the critical value at the 5\% significance level to assess the presence of unit roots in these time series. All the variables become stationary after taking the first difference.

\section{Empirical Results}

To analyze the determinants of the short-term real effective exchange rate (REER) in Uruguay, we initially employed an Ordinary Least Squares (OLS) regression model. The model specification is as follows:

\[
\text{REER} = \beta_0 + \beta_1 \times \text{USLR} + \beta_2 \times \text{M2} + \beta_3 \times \text{CPI} + \beta_4 \times \text{WIR} + \epsilon
\]

However, upon investigation, it was observed that the model exhibited signs of heteroskedasticity, which could potentially lead to inefficient and unreliable standard error estimates. To address this issue, we employed the Newey-West estimator as a robustness check. The error terms $\epsilon$ exhibit autocorrelation and heteroskedasticity, meaning their variances are not constant across time and they might be correlated with each other over different time lags. The Newey-West procedure modifies the calculation of the covariance matrix of the OLS coefficient estimates to address these issues. The formula for the Newey-West corrected covariance matrix ($V$) is

\[
V = \frac{1}{T} \left( X'X \right)^{-1} \left( \sum_{t=1}^T X_t'X_t \hat{\epsilon}_t^2 + \sum_{l=1}^L w_l \sum_{t=l+1}^T \left( X_t'X_{t-l} \hat{\epsilon}_t \hat{\epsilon}_{t-l} + X_{t-l}'X_t \hat{\epsilon}_t \hat{\epsilon}_{t-l} \right) \right) \left( X'X \right)^{-1}
\]

Where:

\begin{itemize}
    \item $T$ is the number of observations.
    \item $X_t$ is the matrix of independent variables at time $t$.
    \item $\hat{\epsilon}_t$ is the residual at time $t$ from the OLS regression.
    \item $L$ is the lag length, which can be determined using various criteria.
    \item $w_l$ is a weight assigned to the lagged terms, which decreases as the lag length increases.
\end{itemize}

The square root of the diagonal elements of the Newey-West adjusted covariance matrix gives the robust standard errors of the OLS coefficient estimates. These standard errors are used to perform hypothesis tests and construct confidence intervals that are more reliable in the presence of heteroskedasticity and autocorrelation.

\begin{table}[H]
\centering
\caption{Regression Results with Newey-West Standard Errors}
\label{tab:model_diff_results}
\begin{tabular}{lcccc}
\hline
\textbf{Variable} & \textbf{Estimate} & \textbf{Std. Error} & \textbf{t value} & \textbf{$Pr(>|t|)$} \\
\hline
(Intercept)   & 1.41617    & 0.31175    & 4.5426  & 2.117e-05 *** \\
USLR          & -1.80743   & 0.71436    & -2.5301 & 0.01353 ***  \\
M2            & -43.92337  & 10.62381   & -4.1344 & 9.284e-05 *** \\
CPI          & -0.61448   & 0.29192    & -2.1049 & 0.03869 ***  \\
WIR           & 1.06235    & 1.34999    & 0.7869  & 0.43383    \\
\hline
\multicolumn{5}{l}{\textbf{Dependent Variable:} REER (Differenced)} \\
\multicolumn{5}{l}{\textbf{Method:} Least Squares} \\
\multicolumn{5}{l}{\textbf{Sample:} 2001.Q4 – 2021.Q3} \\
\multicolumn{5}{l}{\textbf{Included observations:} 79 after adjustments} \\
\multicolumn{5}{l}{\textbf{Newey-West HAC Standard Errors \& Covariance (lag truncation=4)}} \\
\hline
\end{tabular}
\end{table}

In our analysis, we have used the first differences of the variables, namely REER, USLR, M2, CPI, and WIR. The rationale behind this approach stems from the presence of unit roots in the level variables, identified through the Augmented Dickey-Fuller test. Using variables in their levels in the presence of unit roots can lead to spurious regression results, where the statistical significance of the relationship between variables might be misleadingly overstated \citep{hill2018}. Differencing the data helps in transforming the non-stationary time series into stationary ones, thereby eliminating the risk of spurious correlation, and ensuring more reliable and interpretable regression results. In addition to addressing heteroskedasticity and autocorrelation in the regression model using Newey-West standard errors, we also conducted a Shapiro-Wilk normality test on the residuals of the model to check for the normality assumption.

For our model, the test yielded a W-statistic of 0.97432 with a p-value of 0.1116. This result indicates that the null hypothesis of normal distribution in the residuals cannot be rejected at 5\% significance levels. The absence of significant deviations from normality in the residuals suggests that the OLS estimator’s assumptions are reasonably met, lending further credibility to the regression results. This normality check, along with the use of Newey-West standard errors, ensures the robustness and reliability of our findings.

Guided by Hsing’s \citep{hsing2006} analysis of the determinants of the Real Effective Exchange Rate (REER) in Venezuela, we chose similar variables and employed a comparable econometric methodology, specifically adopting the Newey-West procedure. This approach yielded results with coefficient signs that aligned closely with their findings. The study’s findings align well with the Mundell-Fleming model under the IS-LM framework, particularly in the context of an open economy with floating exchange rates like Uruguay.

An increase in the US lending rate (USLR) can lead to capital outflows from Uruguay, as investors seek higher returns, leading to a depreciation of the Uruguayan Peso. This depreciation makes Uruguayan goods more competitive abroad, stimulating exports, and shifting the IS curve to the right, indicating an increase in aggregate demand.

Similarly, an increase in the money supply (M2) would shift the LM curve to the right, as more money in the economy lowers interest rates, stimulating investment and consumption. However, this can also contribute to currency depreciation and inflation (CPI). In the short term, this depreciation further stimulates exports, moving the IS curve rightward. However, high inflation can erode international competitiveness, potentially shifting the IS curve back to the long term if not managed properly. The study’s observation that the world interest rate (WIR) has no significant impact on the REER could be due to Uruguay’s small open economy characteristics, where domestic monetary policy and external factors like USLR play a more dominant role. This underlines the importance of using monetary policy to manage M2 growth and implementing strict measures for inflation control. Adjusting fiscal policies in response to US interest rate fluctuations and developing strategies to enhance export competitiveness also become crucial in this framework, as they can influence the position and movement of the IS and LM curves, impacting the REER.

\section{Conclusion}

In conclusion, this study contributes to the understanding of Uruguay's real effective exchange rate (REER) dynamics within the context of the Mundell-Fleming model. By analyzing key economic indicators and employing econometric techniques, we have uncovered important insights into the determinants of Uruguay's REER.

Our findings indicate that factors such as the US lending rate (USLR), domestic money supply (M2), and inflation (CPI) significantly influence the short-term fluctuations of Uruguay's REER. Specifically, increases in the USLR and M2 lead to REER depreciation, while inflation exerts a negative impact. On the other hand, the world interest rate (WIR) does not show a significant effect on the REER, highlighting the dominance of domestic monetary policy and regional economic factors.

From a policy perspective, our analysis suggests the importance of implementing measures to control money supply growth, manage inflation rates, and respond effectively to changes in external economic conditions, particularly the US lending rate. Tightening monetary policy, adopting inflation control measures, and fostering export competitiveness emerge as crucial strategies for stabilizing the REER and promoting economic stability and growth in Uruguay.

Overall, this study underscores the relevance of the Mundell-Fleming model in understanding the dynamics of exchange rates in open economies like Uruguay. By providing empirical evidence and policy implications, our research aims to inform policymakers and stakeholders in formulating effective monetary and fiscal policies to achieve desired economic outcomes in Uruguay.

\end{document}